\documentclass[12pt]{article}
\usepackage{a4wide}

\usepackage[dvipdfmx,hiresbb]{graphicx} 
\usepackage{mediabb}                    

\usepackage{amsmath,amssymb}
\usepackage{color}
\usepackage[bookmarks,bookmarksnumbered]{hyperref}
\usepackage{cellspace}
\usepackage{tikz}
\usepackage{mathtools}
\usepackage{mathrsfs}
\usetikzlibrary{fadings}
\usetikzlibrary{patterns}
\usetikzlibrary{shadows.blur}
\usetikzlibrary{shapes}
\newcommand{\aln}[1]{\begin{align}#1\end{align}}
\newcommand{\nn}{\nonumber\\}

\begin{document}
\title{\vbox{
\baselineskip 14pt
\hfill \hbox{\normalsize 
}}  \vskip 1cm
\bf \Large Note on general functional flows
\\
 in equilibrium  systems
\vskip 0.5cm
}
\author{
Kiyoharu Kawana,\thanks{E-mail: \tt kkiyoharu@kias.re.kr}
\bigskip\\
\normalsize
\it $^\dagger$ School of Physics, Korean Institute for Advanced Study, Seoul 02455, Korea
\smallskip
}
\date{\today}

\maketitle   
\begin{abstract} 
We study the response of generating functionals to a  variation of parameters (couplings) in equilibrium systems i.e. in quantum field theory (QFT) and equilibrium statistical mechanics.  
These parameters can be either physical ones such as coupling constants or artificial ones which are intentionally introduced such as the renormalization scale in field theories. 
We first derive general functional flow equations for the generating functional (grand-canonical potential) $W[J]$ of the connected diagrams.   
Then, we obtain functional flow equations for the one-particle irreducible ($1$PI) vertex functional (canonical potential) $\Gamma[\phi]$ by performing the Legendre transformation. 
By taking the functional derivatives of the flow equations, we can obtain an infinite hierarchical equations for the $1$PI vertices. 
We also point out that a Callan-Symanzik type equation holds among the vertices when partition function is invariant 
under some changes of the parameters.    
After discussing general aspects of parameter response, we apply our formalism to several examples and reproduce the well-known functional flow equations.  
%
%
%
%
%
%
Our response theory provides us a systematic and general  way to obtain various functional flow equations in equilibrium systems.    

\

%
%
%
%
%
%
%
%

\end{abstract} 

\setcounter{page}{1} 

\newpage  

\tableofcontents   


\section{Introduction}\label{Sec:intro}
For the past half century, Renormalization Group (RG) has played a central role in fundamental physics.  
The essence of RG is the coarse-graining of the microscopic  degrees of freedom and the construction of effective theory at low-energy scales (long distances).  
%
%
In particular, one of the great successes is the explanation of universality class of critical phenomena in nature.   
For example, in the Wilsonian RG~\cite{Wilson:1971bg,Wilson:1971dh,Wilson:1973jj}, we can discuss the RG flow of the effective action by  integrating out the high-energy modes.  
Then, critical behaviors of a system are explained by the (linearlized) flow structures around a fixed point.   
%
%
%
Not only are they very powerful for describing critical phenomena, the RG is also useful for the (non)perturbative calculations in field theories; 
The independence of correlation functions or effective action on the renormalization scale implies nontrivial differential flow equations (Callan-Symanzik equations)~\cite{Callan:1970yg,Symanzik:1970rt}, and  we can systematically improve the perturbative calculations by using these equations~\cite{Bando:1992wy,Iso:2018aoa,Manohar:2020nzp}.

Functional Renormalization Group (FRG)~\cite{Wegner:1972ih,Polchinski:1983gv,Wetterich:1989xg,Fisher:1998kv,Berges:2000ew,Dupuis:2008vr,Bagnuls:2000ae,Dupuis:2020fhh} provides us a non-perturbative way to implement the idea of RG in general many-body systems. 
Essentially, the FRG combines the functional method in mathmatical physics  with the Wilsonian RG by introducing an artificial (IR) cut-off parameter $k$ in the action or Hamiltonian, and we see the response of a system to the continuous change of $k$.   
By construction, the flow equation of the (average)  effective action $\Gamma_k^{}[\phi]$ is one-loop exact, that is, it is expressed as an one-loop integral with the exact two-point correlation function.   
Thus, some reasonable approximations are necessary when one wants to solve the flow equation practically.   
See Refs.~\cite{Bagnuls:2000ae,Dupuis:2020fhh} and references therein for various approximation methods  and for countless applications of the FRG. 
%
%

In this paper, we discuss the basic ideas and formulations that underlie these functional flow approaches in general equilibrium systems i.e. in quantum field theory (QFT) and in equilibrium statistical mechanics.    
The essence of the flow approach is to see the response of a system to the variation of parameters in the action or Hamiltonian. 
%
The parameters can be either physical ones such as coupling constants $\{g_k^{}\}$, temperature $T$, chemical potential $\mu$, or artificial ones that are intentionally added such as the renormalization scale $M$.    
We derive general functional flow equations for the generating functionals in both grand-canonical and canonical formulations.  
In QFT, the grand-canonical (canonical) generating functional $W[J]~(\Gamma[\phi])$ corresponds to the generating functional of the connected diagrams (one-particle irreducible ($1$PI) diagrams). 
In particular, when the partition function is invariant 
under some changes of parameters, 
we can obtain another non-trivial relation for the correlation functions or $1$PI vertices, which can be interpreted as a generalized Callan-Symanzik equation. 
%

After discussing general aspects of parameter response   in Section~\ref{sec:response}, we apply them to several examples and reproduce the well-known flow equations based on our general results. 
%
%
The first (trivial) example is the FRG in QFT, where an IR regulator function $R_k^{}(x-y)$ is introduced in the propagator and we reproduce the Wetterich's equation. 
%
The second example is quantum spin systems (quantum Heisenberg models)~\cite{tasaki2020physics} and it is shown that the canonical generating functional follows the Wetterich's equation by introducing an artificial parameter $t$ and modifying the exchange coupling as  $J_{ij}^{}\rightarrow J_{ij}^{}(t)~,~J_{ij}^{}(t=1)=J_{ij}^{}$.  

The third example is the classical liquid systems~\cite{RevModPhys.48.587,hansen90a}.  
Here, we study two different flow approaches; One is called the ``Hierarchical Reference Theory" (HRT)~\cite{LReatto_1990,Parola_2012,Caillol_2006,doi:10.1080/00268976.2011.621455,Yokota:2021ute} where an artificial IR cut off $k$ is introduced to the microscopic two-body pair potential $v(x-y)$ as in the FRG. 
We show that the resultant flow equation is completely the same as the Wetterich's equation by identifying the cut-off dependent pair potential $v_k^{}(x-y)$ as a regulator $R_k^{}(x-y)$. 
Our general theory also enables us to obtain an exact flow equation in the presence of higher-order many-body potentials $v_n^{}(\{x_i^{}\})~,~n\geq 3$.  
The other application in classical liquids is the density renormalization group which was investigated in our  previous paper~\cite{Iso:2018rag}. 
In this case, we consider a scale transformation  $v(x)~\rightarrow~v(\lambda x)$ and regard $\lambda$ as a response parameter.  
In the previous study, we have relied on the field theoretical technique~\cite{HUBBARD1972245} to derive the flow equations, but they can be derived more directly (and easily) by using our general results of response theory. 

As a last example, we discuss classical nonequilibrium systems, in particular, the Langevin stochastic systems. 
These systems can be treated as equilibrium ones by   casting them into the path-integral forms.    
We again see that the flow equation of the effective action of these nonequlibirum systems is given by the Wetterich's equation when a cut-off regulator $R_k^{}(x-y)$ is added to the propagator.     

In this paper, we mainly focus on theoretical formulations  and general aspects of response theory. 
More detailed analysis in some concrete models/systems is left for future investigation.  


\

The organization of this paper is as follows. 
In Section~\ref{sec:response}, we formulate general response theory.  
We derive the exact functional flow equations for the generating functionals by considering general variations of parameters. 
We also point out that a non-trivial functional relation among the correlation functions or $1$PI vertices can hold when a system has a redundancy under some changes of the parameters.     
In Section~\ref{sec:example}, we apply our general results to typical equilibrium systems and reproduce the well-known flow equations.  
Summary and discussion is presented in Section~\ref{sec:summary}.  

Throughout the paper, we work in the $d$ dimensional flat spacetime with the metric convention $\eta_{\mu\nu}^{}=(-,+,\cdots,+)$.  
%

\section{General response theory}\label{sec:response}
We discuss general parameter response in equilibrium systems. 
The main goal is to derive the exact functional flow equations for the $1$PI generating functional  Eq.~(\ref{general flow 2}). 
%

\subsection{General flow equations in grand-canonical formulation}
We consider a broad class of equilibrium systems whose bare  action $S^{}$ or Hamiltonian $H^{}$ in the $d$ dimensional flat spacetime is given by
\aln{\label{general action}
S^{}~[\phi,\{\lambda_k^{}\};J]~=\sum_{n=1}^\infty \int \left(\prod_{i=1}^n d^{d}x_i\right)\frac{v_n^{}(\{x_i^{}\};\{\lambda_k^{}\})}{n!}\left(\prod_{j=1}^n\phi(x_j^{})\right)+(i\hbar^{-1})^{-1}\int d^{d}x J(x)\phi(x)~, 
\\
\label{general hamiltonian}
H^{}[\phi,\{\lambda_k^{}\};J]~=\sum_{n=1}^\infty \int \left(\prod_{i=1}^n d^{d-1}x_i\right)\frac{v_n^{}(\{x_i^{}\};\{\lambda_k^{}\})}{n!}\left(\prod_{j=1}^n\phi(x_j^{})\right)-\beta^{-1}\int d^{d-1}x J(x)\phi(x)~, 
}
where $x_i^{}=\overrightarrow{x}_i^{}$ denotes a position of the $i$-th particle, 
 $J(x)$ is an external source, 
$\{x_i^{}\}=\{x_i^{}\}_{i=1}^n$, and $\{\lambda_k^{}\}$ represents a set of general parameters (couplings) that can include artificial ones as well.   
%
Here, we simply consider one real scalar model in a continuous spacetime, but generalization to vector models and lattice systems is straightforward.  
See Sections~\ref{sec:quantum spin} and \ref{nonequilibirum systems} for examples. 
%
Note also that 
the microscopic $n$-body potential $v_n^{}(\{x_i^{}\};\{\lambda_k^{}\})$ is symmetric under the permutation of positions and can contain the spacetime derivatives $\partial_\mu^{}$ in general. 
%
%
%
For example, an ordinary scalar QFT corresponds to 
\aln{\quad v_2^{}(x_1^{},x_2^{})=\delta^{(d)}(x_1^{}-x_2^{})(\partial^2-m^2)~,\quad v_n^{}(\{x_i^{}\};\{\lambda_k^{}\})=-\lambda_n^{}\prod_{i=2}^n\delta^{(d)}(x_1^{}-x_i^{})~(\text{for $n\neq 2$})~,
} 
and the two-body classical liquid system corresponds to~\cite{RevModPhys.48.587,hansen90a}
\aln{
v_1^{}(x)=-\frac{1}{2}v(x,x)~,\quad v_2^{}(x,y)=v(x_1^{},x_2^{})~,\quad v_n^{}=0\quad (\text{for $n\geq 3$})~,
} 
where 
$v(x,y)$ is a microscopic two-body pair potential. 
In this case, $\phi(x)=\rho(x)$ corresponds to the density of the system and $J(x)=\beta U(x)$ corresponds to the external chemical potential.  
See Section~\ref{sec:classical liquids} for more details.
To simplify the expressions, we also use the following notations for the first two interactions:
\aln{
\int d^{\tilde{d}}x v_1^{}(x)\phi(x)=\langle v_1^{}|\phi\rangle~,\quad \int d^{\tilde{d}}x\int d^{\tilde{d}}y \phi(x)v_2^{}(x,y)\phi(y)=\langle \phi|v_2^{}|\phi\rangle~.
} 
where $\tilde{d}=d~(d-1)$ for QFT (statistical mechanics).  

The partition function is defined by
\aln{\label{general partition function}
Z[\{\lambda_k^{}\};J]=e^{sW[\{\lambda_k^{}\};J]}&=\begin{dcases}
\int {\cal D}\phi e^{i\hbar^{-1} S^{}[\phi,\{\lambda_k^{}\};J]}
\\
\text{Tr}\left(e^{-\beta H^{}[\phi,\{\lambda_k^{}\};J]}\right)
\end{dcases}~
}
where $s=i\hbar^{-1}~(-\beta)$ for $S^{}~(H^{})$.   
Here, $W[\{\lambda_k^{}\};J]$ corresponds to the generating functional of the connected diagrams (grand-canonical potential) in QFT (statistical mechanics). 

We consider the response of the generating functional $sW[\{\lambda_k^{}\};J]$ to a small change of the parameters $\{\delta \lambda_k^{}\}$:
\aln{
\delta (sW[\{\lambda_k^{}\};J])
&=s\sum_{n=1}^\infty \frac{1}{n!}\int \left(\prod_{i=1}^nd^{\tilde{d}}x_i^{}\right)\delta v_n^{}(\{x_i^{}\};\{\lambda_k^{}\})\left\langle T\prod_{j=1}^n\phi(x_i^{})\right\rangle_J^{}~,
\label{variation of W}
}
where
\aln{\left\langle T\prod_{j=1}^n\phi(x_i^{})\right\rangle_J^{}&=\frac{1}{Z}\begin{dcases} \int {\cal D}\phi \left(\prod_{i=1}^n\phi(x_i^{})\right)e^{i\hbar^{-1}S^{}[\phi,\{\lambda_k^{}\};J]}
\\
\text{Tr}\left(\prod_{i=1}^n\phi(x_i^{})e^{-\beta H^{}[\phi,\{\lambda_k^{}\};J]}\right)
\end{dcases}
\\
&:=G^{(n)}(\{x_i^{}\};\{\lambda_k^{}\};J)~.
}
Then, Eq.~(\ref{variation of W}) can be written as 
\aln{
\frac{d(sW[\{\lambda_k^{}\};J])}{d\lambda_k^{}}
&=s\sum_{n=1}^\infty \frac{1}{n!} \int \left(\prod_{i=1}^n d^{\tilde{d}}x_i^{}\right)
\frac{\partial v_n^{}(\{x_i^{}\};\{\lambda_k^{}\})}{\partial \lambda_k^{}}G^{(n)}(\{x_i^{}\};\{\lambda_k^{}\};J)\quad (\text{for each  $k$})~.
\label{general flow 1}
}
More generally, for a parametric variation $ \lambda_k^{}\rightarrow \lambda_k^{}(t)$, we have
\aln{
\frac{d(sW[\{\lambda_k^{}(t)\};J])}{dt}&=\sum_{k=1}^\infty \frac{d\lambda_k^{}(t)}{dt}\frac{\partial (sW[\{\lambda_k^{}\};J])}{\partial \lambda_k^{}} \bigg|_{\lambda_k^{}=\lambda_k^{}(t)}
\\
&=s\sum_{k=1}^{\infty}\frac{d\lambda_k^{}(t)}{dt}\sum_{n=1}^\infty \frac{1}{n!} \int \left(\prod_{i=1}^n d^{\tilde{d}}x_i^{}\right)
\frac{\partial v_n^{}}{\partial \lambda_k^{}}G^{(n)}(\{x_i^{}\};\{\lambda_k^{}\};J) \bigg|_{\lambda_k^{}=\lambda_k^{}(t)}~.
}
Note that $G^{(n)}$ contains non-connected diagrams and is related to the connected correlation functions defined by
\aln{
\frac{\delta^n (sW[\{\lambda_k^{}\};J])}{\delta J(x_1^{})\cdots \delta J(x_n^{})}&=F^{(n)}(\{x_i^{}\};\{\lambda_k^{}\};J)~.  
}
For example,
\aln{
\label{F1 and G1}
F^{(1)}(x;\{\lambda_k^{}\};J)&=G^{(1)}(x;\{\lambda_k^{}\};J)~,
\\
F^{(2)}(x,y;\{\lambda_k^{}\};J)&=G^{(2)}(x,y;\{\lambda_k^{}\};J)-F^{(1)}(x;\{\lambda_k^{}\};J)F^{(1)}(y;\{\lambda_k^{}\};J)~.\label{F2 and G2}
}
By definition, the connected correlation functions satisfy 
\aln{
\label{F relation}
\frac{\delta F^{(n)}(\{x_i^{}\};\{\lambda_k^{}\};J)}{\delta J(x_{n+1}^{})}&=F^{(n+1)}(\{x_i^{}\};\{\lambda_k^{}\};J)~.
}
For simplicity, we denote the correlation function without the source term as 
\aln{F^{(n)}(\{x_i^{}\};\{\lambda_k^{}\})=F^{(n)}(\{x_i^{}\};\{\lambda_k^{}\};J=0)~.
}
 Eq.~(\ref{general flow 1}) is a general functional flow equation in the grand-canonical formulation. 
 By taking the functional derivatives of Eq.~(\ref{general flow 1}) with respect to $J(x)$ (and putting $J(x)=0$), we can obtain hierarchical equations for the correlation functions.   
 For example, the first two equations are
 \aln{
 \frac{d}{d\lambda_k^{}}F^{(1)}(x;\{\lambda_k^{}\})=s\sum_{n=1}^\infty \frac{1}{n!} \int \left(\prod_{i=1}^n d^{\tilde{d}}x_i^{}\right)
 \frac{\partial v_n^{}}{\partial \lambda_k^{}}\frac{\delta G^{(n)}(\{x_i^{}\};\{\lambda_k^{}\})}{\delta J(x)}~,\label{flow of F1}
 \\
  \frac{d}{d\lambda_k^{}}F^{(2)}(x,y;\{\lambda_k^{}\})=s\sum_{n=1}^\infty \frac{1}{n!} \int \left(\prod_{i=1}^n d^{\tilde{d}}x_i^{}\right)
  \frac{\partial v_n^{}}{\partial \lambda_k^{}}\frac{\delta^2 G^{(n)}(\{x_i^{}\};\{\lambda_k^{}\})}{\delta J(x)\delta J(y)}~.\label{flow of F2}
 }
 The calculation of the right hand sides needs more information of a specific model.  
 %
 In particular, when we consider parameter changes such that only $G^{(1)}$ or  $G^{(2)}$ appears in the right hand side in Eq.~(\ref{general flow 1}), Eqs.~(\ref{F1 and G1})(\ref{F2 and G2})(\ref{F relation}) are  sufficient to determine the flow equations for the higher-order correlation functions $F^{(l)}$.  
 For example, when only $G^{(2)}$ appears in Eq.~(\ref{general flow 1}), the first two equations (\ref{flow of F1})(\ref{flow of F2}) become 
 \aln{\frac{d}{d\lambda_k^{}}F^{(1)}(x)&=\frac{s}{2}
 \int d^{\tilde{d}}x_1^{}\int d^{\tilde{d}}x_2^{}\frac{\partial v_2^{}(x_1^{},x_2^{})}{\partial \lambda_k^{}}
\left\{
F^{(3)}(x,x_1^{},x_2^{})+2F^{(2)}(x,x_1^{})F^{(1)}(x_2^{})
\right\}~,
 \\
 \frac{d}{d\lambda_k^{}}F^{(2)}(x,y)&=\frac{s}{2}
 \int d^{\tilde{d}}x_1^{}\int d^{\tilde{d}}x_2^{}\frac{\partial v_2^{}(x_1^{},x_2^{})}{\partial \lambda_k^{}}
 \nn
&\hspace{2cm}\times \bigg\{
F^{(4)}(x,y,x_1^{},x_2^{})+2F^{(3)}(x,y,x_1^{})F^{(1)}(x_2^{})
+2F^{(2)}(x,x_1^{})F^{(2)}(y,x_2^{})\bigg\}~,
 }
 where we have omitted $\{\lambda_k^{}\}$ in the correlation functions for simplicity. 
 In this case, the flow equations for the higher-order correlation functions $F^{(l)}~(l\geq 3)$ can be  obtained straightforwardly by taking the functional derivative of Eq.~(\ref{flow of F2}) with the use of Eq.~(\ref{F relation}). 
 %

\subsection{General flow equations in canonical formulation}\label{canonical flow}
Usually it is more convenient to consider the generating functional of $1$PI vertices (canonical potential) $\Gamma[\phi]$ instead of the grand-canonical one $W[J]$. 
Here, we derive the flow equation of $\Gamma[\phi]$.

The effective action is defined by the Legendre transformation of $W[\{x_i^{}\};\{\lambda_k^{}\};J]$:
\aln{
\label{effective action}
s\Gamma[\{\lambda_k^{}\};\phi]&=\underset{J}{{\rm min}}\left(sW[\{\lambda_k^{}\};J]-\langle J|\phi\rangle\right)
\\
&=sW[\{\lambda_k^{}\};J_\phi^{}]-\langle J_\phi^{}|\phi\rangle~,
}
where $J_\phi^{}$ is a solution of 
\aln{
\frac{\delta (sW[\{\lambda_k^{}\};J^{}])}{\delta J(x)}=F^{(1)}(x;\{\lambda_k^{}\};J)=\phi(x)~.
\label{Jphi}
}
This implies that the two-point function is given by
\aln{
F^{(2)}(x,y;\{\lambda_k^{}\};J_\phi^{})=\frac{\delta (sW[\{\lambda_k^{}\};J^{}])}{\delta J(x)\delta J(y)}\bigg|_{J=J_\phi^{}}^{}=\frac{\delta \phi(x)}{\delta J_\phi^{}(y)}~
}
for a given field $\phi(x)$. 
%
The parameter derivative is calculated as 
\aln{\frac{d(s\Gamma[\{\lambda_k^{}\};\phi])}{d\lambda_k^{}}&=\frac{d}{d\lambda_k^{}}\left(sW[\{\lambda_k^{}\};J_\phi^{}]-\langle J_\phi^{}|\phi\rangle\right)
\nn
&=\frac{d(sW[\{\lambda_k^{}\};J^{}])}{d\lambda_k^{}}\bigg|_{J=J_\phi^{}}-\int d^{\tilde{d}}x\frac{\partial J_\phi^{}}{\partial \lambda_k^{}}\frac{\delta }{\delta J_\phi^{}(x)} \left(sW[\{\lambda_k^{}\};J_\phi^{}]-\langle J_\phi^{}|\phi\rangle\right)
\nn
&=\frac{d(sW[\{\lambda_k^{}\};J^{}])}{d\lambda_k^{}}\bigg|_{J=J_\phi^{}}~,
}
where the second term in the second line vanishes by Eq.~(\ref{Jphi}). 
Thus, by Eq.~(\ref{general flow 1}), we obtain a general flow equation of $\Gamma[\{\lambda_k^{}\};\phi]$:
\aln{
\label{general flow 2}
\frac{d(s\Gamma[\{\lambda_k^{}\};\phi])}{d\lambda_k^{}}=s\sum_{n=1}^\infty \frac{1}{n!} \int \left(\prod_{i=1}^n d^{\tilde{d}}x_i^{}\right)
\frac{\partial v_n^{}(\{x_i^{}\};\{\lambda_k^{}\})}{\partial \lambda_k^{}}G^{(n)}(\{x_i^{}\};\{\lambda_k^{}\};J_\phi^{})~.
}
%
To derive the flow equations for the higher-oder vertices, we define the $n-$point ($1$PI) vertex by
\aln{c^{(n)}(\{x_i^{}\};\{\lambda_k^{}\};\phi)=\frac{\delta^n (s\Gamma[\{\lambda_k^{}\};\phi])}{\delta \phi(x_1^{})\cdots \delta \phi(x_n^{})}~.
}
In particular, the first two vertices can be written as
\aln{
c^{(1)}(x;\{\lambda_k^{}\};\phi)=-J_\phi^{}(x)~,\quad c^{(2)}(x,y;\{\lambda_k^{}\};\phi)=-\frac{\delta J_\phi^{}(x)}{\delta \phi(y)}~.
\label{c1 and c2}
}
One can check that these vertices and $F^{(2)}$ satisfy
\aln{
&\int d^{\tilde{d}}zF^{(2)}(x,z;\{\lambda_k^{}\};J_\phi^{})c^{(2)}(z,y;\{\lambda_k^{}\};\phi^{})=-\delta^{(\tilde{d})}(x-y)~,\label{relation 1}
\\
&\frac{\delta F^{(2)}(x_1^{},x_2^{};\{\lambda_k^{}\};\phi)}{\delta \phi(x_{3}^{})}=\int d^{\tilde{d}}z\int d^{\tilde{d}}z'F^{(2)}(x_1^{},z;\{\lambda_k^{}\};\phi)c^{(3)}(x_3^{},z,z';\{\lambda_k^{}\};\phi)F^{(2)}(x_2^{},z';\{\lambda_k^{}\};\phi)\label{relation 2}~,
\\
&\frac{\delta c^{(n)}(\{x_i^{}\};\{\lambda_k^{}\};\phi)}{\delta \phi(x_{n+1}^{})}=c^{(n+1)}(\{x_i^{}\};\{\lambda_k^{}\};\phi)\quad (\text{for $n\geq 3$})~.\label{relation 3}
}
To rewrite the flow equation~(\ref{general flow 2}) in terms of the $1$PI vertices, we have to express the $n-$point correlation function $G^{(n)}$ or $F^{(n)}$ by $\{c^{(l)}\}$ in general.  
Then, by taking the functional derivatives of Eq.~(\ref{general flow 2}) with respect to $\phi(x)$, we can obtain the flow equations for the vertices  $\{c^{(l)}\}$.  

In particular, as in the grand-canonical case,  Eqs.~(\ref{relation 1})(\ref{relation 2})(\ref{relation 3}) are sufficient to derive the flow equations for the higher-order vertices when only $G^{(2)}$ appears in Eq.~(\ref{general flow 2}). 
Here, we summarize the first two equations in this case:
\aln{\frac{d}{d\lambda_k^{}}c^{(1)}(x_1^{})&=\frac{s}{2}
\int d^{\tilde{d}}y^{}\int d^{\tilde{d}}y'\frac{\partial v_2^{}(y,y')}{\partial \lambda_k^{}}
  \bigg\{\int d^{\tilde{d}}z\int d^{\tilde{d}}z'
F^{(2)}(y,z)c^{(3)}(x_1^{},z,z')F^{(2)}(z',y')
\nn
&\hspace{3cm}+2\delta^{(\tilde{d})}(x_1^{}-y)\phi(y')
\bigg\}~,
 }
\aln{
 \frac{d}{d\lambda_k^{}}&c^{(2)}(x_1^{},x_2^{})=\frac{s}{2}
 \int d^{\tilde{d}}y\int d^{\tilde{d}}y'\frac{\partial v_2^{}(y,y')}{\partial \lambda_k^{}}
\bigg\{
\int d^{\tilde{d}}z\int d^{\tilde{d}}z'F^{(2)}(y,z)c^{(4)}(x_1^{},x_2^{},z,z')F^{(2)}(z',y')
\nn
&+2\int d^{\tilde{d}}z\int d^{\tilde{d}}z'\int d^{\tilde{d}}\omega \int d^{\tilde{d}}\omega 'F^{(2)}(y,z) c^{(3)}(x_1^{},z,\omega)F^{(2)}(\omega,\omega')c^{(3)}(x_2^{},\omega',z')F^{(2)}(z',y')
\nn
&\hspace{3cm}+2\delta^{(\tilde{d})}(y-x_1^{})\delta^{(\tilde{d})}(y'-x_2^{})
\bigg\}~,
 }
where we have omitted the parameter dependences and $\phi(x)$ in the vertices for simplicity.  
 %
 %
 The flow equations for higher-order vertices $c^{(l)}~(l\geq 3)$ can be straightforwardly obtained by using Eqs.~(\ref{relation 2})(\ref{relation 3}). 
%

\subsection{Generalized Callan-Symanzik equations}\label{sec:CS}
Let us consider a situation such that the variation of a parameter $\lambda_0^{}:=t$  can be compensated by the changes of other variables as
\aln{\label{general compensation}
\Gamma_{}^{}[\{t-\delta t,\lambda_k^{},V\};\phi]=\Gamma_{}^{}[\{t,\lambda_k^{}+\delta \lambda_k^{}(\delta t),V+\delta V(\delta t)\};\phi+\delta \phi(\delta t)]~,
}
or
\aln{
\Gamma_{}^{}[\{t,\lambda_k^{},V\};\phi]=\Gamma_{}^{}[\{t+\delta t,\lambda_k^{}+\delta \lambda_k^{}(\delta t),V+\delta V(\delta t)\};\phi+\delta \phi(\delta t)]~,
}
which can be interpreted as a symmetry of the generating functional. 
Note that we have also included the volume dependence explicitly. 
The above relations mean 
\aln{-\frac{d\Gamma_{}^{}[\{t,\lambda_k^{},V\};\phi]}{dt}&=\left(\sum_k\frac{\delta \lambda_k^{}(\delta t)}{\delta t}\frac{\partial}{\partial \lambda_k^{}}+\frac{\delta V(\delta t)}{\delta t}\frac{\partial}{\partial V}+\int  d^{\tilde{d}}x \frac{\delta \phi(\delta t)}{\delta t}\frac{\delta}{\delta \phi(x)}\right)\Gamma_{}^{}[\{t,\lambda_k^{},V\};\phi]
\\
:&={\cal D}_t^{}\Gamma_{}^{}[\{t,\lambda_k^{},V\};\phi]~.
}
By using the general flow equation (\ref{general flow 2}) (with $\lambda_k^{}\rightarrow t$), 
we have another relation among the correlation functions:
\aln{
{\cal D}_t^{}(s\Gamma_{}^{}[\{t,\lambda_k^{}\},V;\phi])
=-s\sum_{n=1}^\infty \frac{1}{n!} \int \left(\prod_{i=1}^n d^{\tilde{d}}x_i^{}\right)\frac{\partial v_{n}^{}(\{x_i^{}\},\{t,\lambda_k^{}\})}{\partial t}
G^{(n)}(\{x_i^{}\};\{t,\lambda_k^{}\};J_\phi^{})~,
}
which 
corresponds to the Callan-Symanzik equation in QFT.~\footnote{In QFT, the effective action is completely independent of the renormalization scale $t=\log M$ 
and $\Gamma_t^{}[\{\lambda_k^{}\};\phi]$ satisfies the Callan-Symanzik equation:
\aln{
0=\frac{d\Gamma_{t}^{}[\{\lambda_k^{}\};\phi]}{dt}=\left(\frac{\partial}{\partial t}+\sum_k \beta_k^{}\frac{\partial}{\partial \lambda_k^{}}-\int d^dx \gamma \frac{\partial}{\partial \log \phi(x)}\right)\Gamma_{t}^{}[\{\lambda_k^{}\};\phi]~. 
\label{CS in QFT}
}
}
In particular, when $t$ is an artificial parameter such that $t=1$ corresponds to the original  system of interest (i.e. $v_n^{}(\{t=1,\lambda_k^{}\})=v_n^{}(\{\lambda_k^{}\})$),  we have
\aln{{\cal D}_t^{}(s\Gamma_{}^{}[\{t,\lambda_k^{}\},V;\phi])\bigg|_{t=1}^{}
=-s\sum_{n=1}^\infty \frac{1}{n!} \int \left(\prod_{i=1}^n d^{\tilde{d}}x_i^{}\right)\frac{\partial v_{n}^{}(\{x_i^{}\},\{t,\lambda_k^{}\})}{\partial t}\bigg|_{t=1}^{}
G^{(n)}(\{x_i^{}\};\{\lambda_k^{}\};J_\phi^{})~,
}
which indicates another nontrivial relation among the correlation functions and vertices in the original system. 

Let us see a couple of examples.    

\

\noindent{\bf GENERAL PRESSURE EQUATION}\\
Here, we represent the generating functional as $W[\{v_n^{}(\{x_i^{}\})\},V;J(x)]$.  
Consider a scale transformation of the volume  $V\rightarrow (1+\tilde{d}\epsilon)V$, where $\epsilon$ is an infinitesimal parameter and regarded as $t$ in the above general discussion.  
In this case, by the definition of the partition function~(\ref{general partition function}), $W$  satisfies 
\aln{W[\{v_n^{}(\{x_i^{}\})\},(1+\tilde{d}\epsilon)V;J(x)]=W[\{v_n^{}(\{(1-\epsilon)x_i^{}\})\},V;J((1-\epsilon)x)]~,
}
which leads to
\aln{
({\rm L.H.S})=&\frac{d(s W[\{v_n^{}(\{x_i^{}\})\},(1+\tilde{d}\epsilon)V;J(x)])}{d\epsilon}\bigg|_{\epsilon=0,J=0}^{}=\tilde{d}\frac{\partial (sW[\{v_n^{}(\{x_i^{}\})\},V])}{\partial \log V}
\nn
({\rm R.H.S})=&-s\sum_{n=1}^\infty \frac{1}{n!} \int \left(\prod_{i=1}^n d^{\tilde{d}}x_i^{}\right)G^{(n)}(\{x_i^{}\})\sum_{i=1}^nx_i^\mu\partial_\mu^{(i)}v_n^{}(\{x_i^{}\})
-\int d^{\tilde{d}}x \int d^{\tilde{d}}yy^\mu\frac{\delta (\partial_\mu^{(y)}J(y))}{\delta J(x)}F^{(1)}(x)
}
\aln{
\therefore~& \frac{\partial (sW[\{v_n^{}(\{x_i^{}\})\},V;J])}{\partial \log V}=\int d^{\tilde d}xF^{(1)}(x)-\frac{s}{\tilde{d}}\sum_{n=1}^\infty \frac{1}{n!} \int \left(\prod_{i=1}^n d^{\tilde{d}}x_i^{}\right)G^{(n)}(\{x_i^{}\})\sum_{i=1}^nx_i^\mu\partial_\mu^{(i)}v_n^{}(\{x_i^{}\})~.
\label{general pressure equation}
}
This is a generalization of the {\it pressure equation} in the classical liquid systems~\cite{RevModPhys.48.587,hansen90a}. 
In fact, the classical liquid systems correspond to
\aln{\frac{\partial (-\beta W[V])}{\partial V}\bigg|_{T,\mu}^{}=\beta p~,\quad F^{(1)}(x)=\rho~,
}
and Eq.~(\ref{general pressure equation}) becomes 
\aln{\frac{p}{T}=\rho+\frac{1}{T(d-1)}\sum_{n=1}^\infty \frac{1}{n!} \int \left(\prod_{i=1}^n d^{d-1}x_i^{}\right)G^{(n)}(\{x_i^{}\})
\sum_{i=1}^nx_i^\mu\partial_\mu^{(i)}v_n^{}(\{x_i^{}\})~.
}
In Section~\ref{sec:classical liquids}, we will see another  example of parameter redundancy in the classical simple  liquid systems.  

\

\noindent{\bf SYSTEMS WITH SCALING POTENTIALS}\\  
When the microscopic potential $v_{n}^{}(\{x_i^{}\},\{t,\lambda_k^{}\})$ satisfies the following scaling property 
\aln{
v_{n}^{}(\{x_i^{}\},\{t,\lambda_k^{}\})
=(t^{\Delta })^nv_n^{}(\{x_i^{}\},\{\lambda_k^{}\})~
}
for some (artificial) parameter $t$, 
the effective action satisfies 
%
\aln{
\Gamma_{}[\{t,\lambda_k^{}\},V;\phi]
=\Gamma_{}^{}[\{\lambda_k^{},V\};t^{\Delta^{}}\phi]~,
}
which leads to 
\aln{
&\int d^{\tilde{d}}x \frac{\delta (s\Gamma_{}^{}[\{\lambda_k^{},V\};\phi])}{\delta \log \phi(x)}=s\sum_{n=1}^\infty \frac{1}{(n-1)!} \int \left(\prod_{i=1}^n d^{\tilde{d}}x_i^{}\right)v_n^{}(\{x_i^{}\},\{\lambda_k^{}\}) G^{(n)}(\{x_i^{}\};\{\lambda_k^{}\};J_\phi^{})
\\
&\therefore ~\int d^{\tilde{d}}x \phi(x)c^{(1)}(x,\{\lambda_k^{}\};\phi)=s\sum_{n=1}^\infty \frac{1}{(n-1)!} \int \left(\prod_{i=1}^n d^{\tilde{d}}x_i^{}\right)v_n^{}(\{x_i^{}\},\{\lambda_k^{}\}) G^{(n)}(\{x_i^{}\};\{\lambda_k^{}\};J_\phi^{})~.
\label{conformal relation} 
}
For $\phi=$constant, this gives another relation between thermodynamic quantities and correlation functions.   

%
For example, we can consider a two-body system with 
\aln{
v_2^{}(x-y)=\frac{v_0^{}}{|x-y|^m}~,\quad v_n^{}(\{x_i^{}\})=0\quad (n\neq 2)~.
}
which satisfies the scaling property $v_2^{}(tx)=t^{-m}v_2^{}(x)$. 
In this case, Eq.~(\ref{conformal relation}) becomes
\aln{\int  d^{\tilde{d}}x \phi(x)c^{(1)}(x)
=s\int d^{\tilde{d}}x\int d^{\tilde{d}}y v_2^{}(x-y) G^{(2)}(x,y;J_\phi^{})~.
}
By putting $\phi(x)=\phi=$constant, 
we then obtain
\aln{-J_\phi^{} V\phi=s\int d^{\tilde{d}}x\int d^{\tilde{d}}y v_2^{}(x-y) G^{(2)}(x,y;J_\phi^{})~,
} 
where we have used Eq.~(\ref{c1 and c2}).
This resembles the sum rules of correlation functions in classical liquid systems~\cite{RevModPhys.48.587,hansen90a}.~\footnote{In liquid systems, $\phi=\rho$ corresponds to the density and $J_\rho^{}$ corresponds the chemical potential $\beta \mu$.  
Thus, $V\phi=N$ is the total particle number in the left hand side. 
}  

%

%

%

\section{Examples}\label{sec:example}
We apply the general results developed in the previous section to several (non)equilibrium systems and reproduce the well-known functional flow equations discussed in the literatures.   
%

\subsection{Functional renormalization group in QFT}\label{sec:FRG}
In the functional renormalization approach in QFT~\cite{Berges:2000ew,Bagnuls:2000ae,Dupuis:2020fhh}, we introduce a cut off $k$ in the two-point vertex as 
\aln{
v_2^{}(x,y;m^2,k)=\delta^{(d)}(x^{}-y^{})(\partial^2-m^2)-R_k^{}(x-y)~,
}
where $R_k^{}(x-y)$ is a regulator function whose Fourier  mode qualitatively satisfies 
\aln{
\tilde{R}_k^{}(p)\sim \begin{cases} 0  & \text{for $p\gg k$}
\\
{\cal O}(k^2) & \text{for $p\ll k$}
\end{cases}~.
}
Namely, it suppresses the low-energy modes and takes only  the high-energy modes in the partition function.  
See Refs.~\cite{Bagnuls:2000ae,Dupuis:2020fhh} and references therein for various proposals of $\tilde{R}_k^{}(p)$.

In this case, the general flow equation~(\ref{general flow 2}) becomes
\aln{\frac{d(s\Gamma[k;\phi])}{dk}&=\frac{s}{2}\int d^dx\int d^dy \partial_k^{}R_k^{}(x-y)G_k^{(2)}(x,y;J_\phi^{})
\\
&=\frac{s}{2}{\rm Tr}\left((\partial_k^{}R_k^{})G_k^{(2)}\right)~,
}
with $s=(i\hbar^{-1})^{-1}$. 
Or by introducing a new effective action and two-point vertex by
\aln{
s\Gamma_k^{}[\phi]:
&=
s\Gamma[k;\phi]+\frac{s}{2}\langle \phi|R_k^{}|\phi\rangle
\\
\Gamma_k^{(2)}(x,y):&=\frac{\delta^2 \Gamma_k^{}[\phi]}{\delta \phi(x)\delta \phi(y)}=s^{-1}c_k^{(2)}(x,y)+R_k^{}(x,y)~,
}
we have
\aln{\frac{d(s\Gamma_k^{}[\phi])}{dk}&=\frac{\partial(s\Gamma[k;\phi^{}])}{\partial k}+\frac{s}{2}\langle \phi|\partial_k^{}R_k^{}|\phi\rangle
\\
&=\frac{s}{2}\int d^dx\int d^dy (\partial_k^{}R_k^{})G_k^{(2)}(x,y;J_\phi^{})
+\frac{s}{2}\langle \phi|\partial_k^{}R_k^{}|\phi\rangle_{J_\phi^{}}
\\
&=\frac{s}{2}\int d^dx\int d^dy (\partial_k^{}R_k^{})F_k^{(2)}(x,y;J_\phi^{})~.
\\
\therefore \quad  \frac{d(i\hbar^{-1}\Gamma_k^{}[\phi])}{dk}&=-\frac{1}{2}\int d^dx\int d^dy (\partial_k^{}R_k^{})(\Gamma_k^{(2)}-R_k^{})^{-1}
=-\frac{1}{2}{\rm Tr}\left[(\partial_k^{}R_k^{})(\Gamma_k^{(2)}-R_k^{})^{-1}\right]~,
\label{Wetterich equation}
}
which is the Wetterich's equation~\cite{Wetterich:1989xg} in the Minkowski spacetime. 
The Euclidean case is obtained by the replacement 
\aln{\Gamma_k^{}[\phi]=i\Gamma_{k}^{E}[\phi]~,\quad \Gamma_k^{(2)}=-{\Gamma_{k}^E}^{(2)}~.
} 
Note that the right-hand side is ${\cal O}(s^0)={\cal O}(\hbar^{0})$, which indicates that this term corresponds to the one-loop diagram. 
As generally explained in Section~\ref{canonical flow},  we can obtain the flow equations for higher-order vertices by taking the functional derivative of Eq.~(\ref{Wetterich equation}) with respect to $\phi(x)$.

\subsection{Quantum spin systems}\label{sec:quantum spin}
Quantum spin systems~\cite{tasaki2020physics} have played  an significant role in statistical mechanics because they capture the essence of quantum many-body systems. 
We can also develop the functional flow in these systems~\cite{Krieg_2019}. 
In the following, a $(d-1)$ dimensional lattice is represented by $\Lambda$, and $i,j,\cdots$ denote the sites. 
Besides, a collection of bonds, i.e. $(i,j)$ such that $i\neq j$, is represented by $\mathscr{B}$, which determines the interactions of sites.

We consider the quantum Heisenberg model:
\aln{
\hat{H}&=\sum_{(i,j)\in \mathscr{B}}J_{ij}^{}\hat{\mathbf{S}}_i^{}\cdot \hat{\mathbf{S}}_j^{}-\beta^{-1}\sum_{i\in \Lambda}\mathbf{U}_i^{}\cdot \hat{\mathbf{S}}_i^{}~,
\label{quantum spin}
\\
Z[\{\mathbf{U}_i^{}\}]&=\exp\left(-\beta W[\{\mathbf{U}^{i}\}]\right)={\rm Tr}(e^{-\beta \hat{H}})~.
}
where $J_{ij}^{}$ is the (anti)ferromagnetic  coupling constant and $\mathbf{H}_{i}=\beta^{-1}\mathbf{U}_{i}^{}=\beta^{-1}(U_{i}^x,U_{i}^{y},U_{i}^z)$ is an external magnetic field. 
%
%
The spin operators $\hat{\mathbf{S}}_i^{}=(\hat{S}^x_i,\hat{S}^y_i,\hat{S}^z_i)$ satisfy the commutation relation
\aln{[\hat{S}^\alpha_i,\hat{S}^\beta_j]=i\delta_{ij}^{}\epsilon^{\alpha \beta \gamma }\hat{S}^\gamma_i~,
}
where $\epsilon^{\alpha \beta \gamma}$ is the totally antisymmetric tensor. 
The magnitude of the spin is defined by $\hat{\mathbf{S}}^2=S(S+1)$. 

Now we introduce an artificial parameter $t\in [0,1]$ and deform  the couplings as $J_{ij}^{}\rightarrow J_{ij}^{}(t)$ with the boundary condition $J_{ij}^{}(t=1)=J_{ij}^{}$. 
Correspondingly, we represent the generating functionals as \aln{W[t;\{\mathbf{U}_i^{}\}]=W_t^{}[\{\mathbf{U}_i^{}\}]~,\quad \Gamma^{}[t;\{\mathbf{S}_i^{}\}]=\Gamma_t^{}[\{\mathbf{S}_i^{}\}]~.
}
As for the boundary (initial) condition at $t=0$, it is often assumed that $J_{ij}^{}(t=0)$ corresponds the coupling of some (exactly) solvable system~\cite{Krieg_2019}. 
The correlation functions are defined by
\aln{G^{i_1^{}\cdots i_n^{}}_{\alpha_1^{}\cdots \alpha_n^{}}(t;\{\mathbf{U}_i^{}\}):&=\frac{1}{Z[t;\{\mathbf{U}_i^{}\}]}\frac{\delta^nZ[t;\{\mathbf{U}_i^{}\}]}{\delta U_{i_1^{}}^{\alpha_1^{}}\cdots \delta U_{i_n^{}}^{\alpha_n^{}}}~,
\\
F^{i_1^{}\cdots i_n^{}}_{\alpha_1^{}\cdots \alpha_n^{}}(t;\{\mathbf{U}_i^{}\}):&=\frac{\delta^n(-\beta W_t^{}[\{\mathbf{U}_i^{}\}])}{\delta U_{i_1^{}}^{\alpha_1^{}}\cdots \delta U_{i_n^{}}^{\alpha_n^{}}}~,
\\
c^{i_1^{}\cdots i_n^{}}_{\alpha_1^{}\cdots \alpha_n^{}}(t;\{\mathbf{U}_i^{}\}):&=\frac{\delta^n(-\beta \Gamma_t^{}[\{\mathbf{S}_i^{}\}])}{\delta S_{i_1^{}}^{\alpha_1^{}}\cdots \delta S_{i_n^{}}^{\alpha_n^{}}}~.
}
Then, by using the general result~(\ref{general flow 2}), we obtain the flow equation for the canonical potential as  
\aln{\frac{d(-\beta \Gamma_t^{}[\{\mathbf{S}_i^{}\}])}{dt}&=-\beta \sum_{\alpha=x,y,z}\sum_{(i,j)\in \mathscr{B}}(\partial_t^{}J_{ij}^{}(t))G^{ij}_{\alpha\alpha}(t;\{\mathbf{U}_i^{S}\})
\\
&=-\beta\sum_{\alpha=x,y,z} \sum_{(i,j)\in \mathscr{B}}(\partial_t^{}J_{ij}^{}(t))\left[F^{ij}_{\alpha\alpha}(\{\mathbf{U}_i^{S}\})+S_{i}^{\alpha}S_j^{\alpha}\right]~.
}
where $\mathbf{U}_i^{S}$ is the solution of the Legendre transformation. 
Note that we do not have a symmetric factor $\frac{1}{2}$ in this case because of the definition of the Hamiltonian~(\ref{quantum spin}). 
As in the FRG in QFT, we can also introduce another canonical functional 
\aln{
-\beta \mathscr{C}_t^{}[\{\mathbf{S}_i^{}\}]:
&=
-\beta \Gamma_t^{}[\{\mathbf{S}_i^{}\}]+\beta \sum_{(i,j)\in \mathscr{B}}{R}_{ij}^{}(t)\mathbf{S}_i^{}\cdot \mathbf{S}_j^{}~.
}
where ${R}_{ij}^{}(t)=J_{ij}^{}(t)-J_{ij}^{}$. 
Now we have
\aln{\frac{d(-\beta \mathscr{C}_t^{}[\{\mathbf{S}_i^{}\}])}{dk}&=-\beta\sum_{\alpha=x,y,z}\sum_{(i,j)\in \mathscr{B}}(\partial_t^{}J_{ij}^{}(t))F^{ij}_{\alpha\alpha}(t;\{\mathbf{U}_i^{S}\})
\\
&=-\sum_{\alpha=x,y,z} \sum_{(i,j)\in \mathscr{B}}(\partial_t^{}R_{ij}^{}(t))\left([\mathbf{\Gamma}^{(2)}_t]^{ij}_{\alpha \alpha}+\partial_t^{}R_{ij}^{}(t)\right)^{-1}
\\
&=-{\rm Tr}\left[(\partial_{t}\mathbf{R})(\mathbf{\Gamma}_t^{(2)}+\mathbf{R}_t^{})^{-1}\right]~,
\label{flow in quantum spin}
}
where
\aln{[\mathbf{\Gamma}^{(2)}_t]^{ij}_{\alpha \beta}=\frac{\delta \mathscr{C}_t^{}[\{\mathbf{S}_i^{}\}]}{\delta S_i^\alpha \delta S_i^\beta}~,\quad [\mathbf{R}_t^{}]^{ij}_{\alpha\beta}=\delta_{\alpha\beta}^{}R_{ij}^{}(t)~.
}
Generalization to higher order interactions such as $J_{ijkl}^{}(t)(\hat{\mathbf{S}}_i^{}\cdot \hat{\mathbf{S}}_j^{})(\hat{\mathbf{S}}_{k}^{}\cdot \hat{\mathbf{S}}_l^{})$ is also straightforward by using the general result~(\ref{general flow 2}).  
%

\subsection{Classical liquids}\label{sec:classical liquids}
Next example is the classical liquid system:
\aln{
H_N^{}=\sum_{i=1}^N\frac{p_i^2}{2m}+V_N^{}(\{x_i^{}\}_{i=1}^N)~,
}
where 
\aln{V_N^{}(\{x_i^{}\}_{i=1}^N)=\sum_{i<j}^Nv(x_i^{},x_j^{})+\sum_{i<j<k}^Nv_3^{}(x_i^{},x_j^{},x_k^{})+\cdots
}
is a general potential energy of $N$ particles. 
%
%
The grand-canonical partition function $\Xi$ and grand potential $W$ are defined by
\aln{
\Xi [T,V;U]&=\exp\left(-\beta W[T,V;U]\right)
\\
 &=\sum_{N=0}^\infty \frac{1
 }{N!}\int_V d^{d-1}x_1^{}\int d^{d-1}p_1^{}\cdots
\int_V d^{d-1}x_N^{}\int d^{d-1}p_N^{}\exp\left(
-\beta H_N^{}+\beta \sum_{i=1}^N U(x_i^{})
\right)
\nn
&=\sum_{N=0}^\infty \frac{(2\pi mT)^{N(d-1)/2}}{N!}\int_V d^{d-1}x_1^{}\cdots
\int_V d^{d-1}x_N^{}\exp\left(
-\beta V_N^{}
+\beta\sum_{i=1}^N U(x_i^{}) 
\right)~,
\label{grand-canonical partition function}
}
where 
$U(x)$ is a position-dependent chemical potential. 
The usual thermodynamic equilibrium corresponds to $U(x)=\mu$.    
%
By using the density operator 
\aln{
\rho(x):=\sum_{i=1}^N\delta^{(d-1)}(x-x_i^{})~,
}
the grand-canonical partition function can be rewritten as
\aln{
\Xi^{}[T,\mu,V;U]&=\sum_{N=0}^\infty \frac{(2\pi mT)^{N(d-1)/2}}{N!}\int_V d^{d-1}x_1^{}\cdots
\int_V d^{d-1}x_N^{}
\nn
&\quad \times \exp\left(
-\frac{1}{2}\langle \rho|\beta v|\rho\rangle
+\frac{\beta }{2}\int d^{d-1}x v(x,x)\rho(x)+\cdots +\langle \beta U | \rho\rangle
\right)~,\label{eq: partition 2}
}
where $\cdots$ represents the higher-order potential terms and 
\aln{
& \langle \rho|\beta v|\rho\rangle= \int d^{d-1}x\int d^{d-1}y\rho(x)\beta v(x,y)\rho(y)~,
\\
& \langle \beta U | \rho\rangle=\beta\int d^{d-1}x U(x) \rho(x)~.
}
Comparing Eq.~(\ref{eq: partition 2}) with general Hamiltonian~(\ref{general hamiltonian}), we can read 
\aln{v_1^{}(x)=-\frac{v(x,x)}{2}~,\quad v_2^{}(x,y)=v(x,y)
} 
in this case.  
%
 
%
The Legendre transformation of $-\beta W[T,V;U]$ with respect to $U(x)$ gives the canonical free energy $-\beta\Gamma [T,V;\rho]$: 
\aln{
-\beta\Gamma^{}[T,V;\rho]&=\underset{U}{\text{Min}}\left(-\beta W^{}[T,V;U]-\langle \beta U|\rho\rangle
\right)
\nn
&=-\beta W^{}[T,V;U_\rho^{}]-\beta \int d^{d-1}xU_\rho^{}(x)\rho(x)~,
\label{Helmholtz free energy}
}
where $\rho(x)$ represents a general density field, and $U_\rho^{}(x)$ is a solution of 
\aln{
\frac{\delta(-\beta W^{}[T,V;U])}{\delta (\beta U(x))}=\rho(x)~.
\label{eq: def of minimum}
}

\

\noindent {\bf HIERARCHICAL REFERENCE THEORY}\\
In the following, we focus on the simple liquids, that is, 
\aln{v(x,y)=v(|x-y|)~,\quad v_{n}^{}(\{x_i^{}\})=0~\text{for $n\geq 3$}~.
}
%
In this case, the Fourier mode 
\aln{\tilde{v}^{}(q)=\int d^{d-1}x e^{-iq\cdot x}v^{}(x)
}
is a function of $|q|$ and real $\tilde{v}^{}(q)^*=\tilde{v}(-q)=\tilde{v}(q)$. 
In the HRT~\cite{LReatto_1990,Parola_2012,Yokota:2021ute},  an artificial IR cut-off $k$ is introduced to the Fourier mode of the two-body pair potential $v(x)$ as
\aln{
\tilde{v}_k^{}(q)\sim \begin{cases}\tilde{v}(q) & \text{for $|q|\gg k$}
\\
0 & \text{for $|q|\ll k$}
\end{cases}~,\label{cut off potential}
}
with the boundary conditions 
\aln{\label{boundary conditions}
\lim_{k\rightarrow \infty}\tilde{v}_k^{}(q)=\tilde{v}_R^{}(q)~,\quad 
\lim_{k\rightarrow 0}\tilde{v}_k^{}(q)=\tilde{v}(q)~,
}
where $v_R^{}(x)$ is the pair potential of some reference system. 
In the literatures~\cite{LReatto_1990,Parola_2012,Caillol_2006,doi:10.1080/00268976.2011.621455}, a repulsive potential (such as the hard core potential) is often chosen for $v_R^{}(x)$.~\footnote{Recently, a new approach using the cavity distribution functions was proposed in Ref.~\cite{Yokota:2021ute}.  
The utilization of the cavity distribution functions eliminates possible divergences coming from the strong short-range repulsive potential.   
}
The expression~(\ref{cut off potential}) is not mathematically complete yet, but we do not need an exact form of $\tilde{v}_k^{}(q)$ to derive the functional flow equation.    
All the quantities calculated by using $v_k^{}(x,y)$ 
are represented as   
\aln{v_{k1}^{}(x)&=-\frac{v_k^{}(x,x)}{2}~,\quad v_{k2}^{}(x,y)=v_k^{}(x,y)~,
\\
W_k^{}[T,V;U]~,&\quad \Gamma_k^{}[T,V;\rho]~,\quad F^{(l)}_k(x_1^{},\cdots,x_l^{})~,\quad  c^{(l)}_k(x_1^{},\cdots,x_l^{})~.
}
Now, according to the general result (\ref{general flow 2}), we have
\aln{\frac{d(-\beta \Gamma_k^{}[T,V;\rho])}{dk}&=-\beta \bigg[\int d^{d-1}x \partial_k^{}v_{k1}^{}(x)F^{(1)}_k(x)
\nn
&\quad \quad +\frac{1}{2}\int d^{d-1}x\int d^{d-1}y \partial_k^{}v_{k2}^{}(x,y)\left(F^{(2)}_k(x,y)+F^{(1)}_k(x)F^{(1)}_k(y)\right)
\bigg]
\nn
=&-\frac{\beta}{2}\int d^{d-1}x\int d^{d-1}y \partial_k^{}v_{k}^{}(x,y)n_k^{(2)}(x,y)~,\label{canonical flow in HRT}
}
where 
\aln{
n_k^{(2)}(x,y)=F^{(2)}_k(x,y)+F^{(1)}_k(x)F_k^{(1)}(y)-\delta^{(d-1)}(x-y)F^{(1)}_k(y)
}
is called the total correlation function in liquid theory. 

To derive a more simple expression, we introduce another canonical free energy ${\cal A}_k^{}$ as follows: 
\aln{
-\beta {\cal{A}}_{k}[T,V;\rho]=-\beta \Gamma_{k}^{}[T,V;\rho]&+\frac{\beta}{2}\int d^{d-1}x\int d^{d-1}y\left\{v_{k}^{}(x,y)-v(x,y)\right\}\{\rho(x)\rho(y)-\delta^{(d)}(x-y)\rho(x)\}~,
\label{def of Ak}
}
which satisfies 
\aln{\lim_{k\rightarrow \infty}{\cal{A}}_{k}[T,V;\rho]=\Gamma_{R}^{}[T,V;\rho]~,\quad \lim_{k\rightarrow 0}{\cal{A}}_{k}[T,V;\rho]=\Gamma_{}^{}[T,V;\rho]~,
}
where $\Gamma_{R}^{}[T,V;\rho]$ is the canonical potential for a reference system.  
By taking the $k$ derivative of $-\beta {\cal A}_k^{}$, we obtain 
\aln{
\frac{d(-\beta {\cal{A}}_k^{}[T,V;\rho])}{d k}
&=-\frac{\beta }{2}\int d^{d-1}x\int d^{d-1}x\partial_k^{}v_{k}^{}(x-y)F_k^{(2)}(x-y)~.
\nn
&=-\frac{\beta }{2}{\rm Tr}[(\partial_k^{}v_k^{})F_k^{(2)}]~,
\label{flow equation in HRT}
}
We represent the two-point vertex defined by $\beta {\cal A}_k^{}[T,V;\rho]$ as 
\aln{
C_k^{(2)}(x-y):=\frac{\delta (\beta {\cal A}_k^{}[T,V;\rho])}{\delta \rho(x)\delta \rho(y)}\bigg|_{\rho(x)=n}^{}=-c_k^{(2)}(x-y)-\beta v_{k}^{}(x-y)+\beta v(x-y)~,
\label{relation of C2}
}
which means that the inverse of $F_k^{(2)}(x)$ is given by $C_k^{(2)}(x)+\beta v_{k}^{}(x)-\beta v(x)$.  
Thus, Eq.~(\ref{flow equation in HRT}) can be also written as
\aln{\frac{d(\beta {\cal{A}}_k^{}[T,V;\rho])}{dk}
&=\frac{1}{2}{\rm Tr}\left[\partial_k^{}(\beta v_{k}^{})(C_k^{(2)}+\beta v_{k}^{}-\beta v^{})^{-1}\right]~,
\label{flow equation 2 in HRT}
}
which is exactly the same form as the Wetterich's equation~(\ref{Wetterich equation}) with the identification $\beta (v_k^{}-v)\rightarrow R_k^{}$. 
This fact indicates that the simple liquid systems  essentially belong to the same universality class of  the scalar QFT.  
See Refs.~\cite{LReatto_1990,Parola_2012} and references therein for the study of critical phenomena based on  the above functional flow equation. 
%
%

\

\noindent {\bf DENSITY RENORMALIZATION GROUP}\\
Another application is the density renormalization group  which was investigated in our previous paper~\cite{Iso:2018rag}. 
We are interested in how the classical liquid systems respond to the variation of the density $\rho(x)=n$.  
To see this, we consider the scale transformation of the pair potential:
\aln{
v(x)\quad \rightarrow \quad v(\lambda x)~,\quad \lambda>0~, 
}
where $\lambda$ is regarded as one of the parameters in the general Hamiltonian~(\ref{general hamiltonian}). 
The corresponding canonical potential is represented as $\Gamma_\lambda^{}[T,V;\rho]$. 
Then, by repeating the same calculation as Eq.~(\ref{canonical flow in HRT}), we obtain the flow equation 
\aln{
\frac{d(-\beta \Gamma_\lambda^{}[T,V;\rho])}{d\lambda}=-\frac{\beta}{2}\int d^{d-1}x\int d^{d-1}y~&n_\lambda^{(2)}(x,y)(x-y)^i\partial_i^{}v(\lambda (x-y))~.
\label{flow in scale}
}
In this case, we can use the idea developed in Section~\ref{sec:CS}, that is, the scale parameter $\lambda$ can be compensated by other variables.  
In fact, the original grand partition function satisfies 
\aln{
&\Xi_{\lambda}^{}[T,V;U(x)]:=\Xi_{}^{}[T,V;U(x)] \bigg|_{v(x)\rightarrow v(\lambda x)}^{}
\nn
=&\sum_{N=0}^\infty \frac{(2\pi mT)^{N(d-1)/2}}{N!}
\int_V \left(\prod_{i=1}^Nd^{d-1}x_i^{}\right)
\exp\left(
-\beta\sum_{i<j}v(\lambda (x_i^{}-x_j^{}))+\beta\sum_i U(x_i^{}) \right)
\nn
=&\sum_{N=0}^\infty \frac{(2\pi mT)^{N(d-1)/2}
}{N!}
\int_{\lambda^{d-1}V}\left(\prod_{i=1}^Nd^{d-1}x_i^{}\right)
\exp\left(
-\beta\sum_{i<j}v(x_i^{}-x_j^{})+\beta\sum_i \left\{U(x_i^{}/\lambda)-(d-1)T\log \lambda\right\}\right)
\nn
=&\Xi_{}^{}[T,\lambda^{d-1}V;U(x/\lambda)-(d-1)T\log \lambda]~.
}
We see that the change of the pair potential under the scale transformation, $v(x) \rightarrow v(\lambda x)$, 
can be compensated by the changes of the chemical potential $\mu$, the volume $V$, and the external chemical potential $U(x)$. 
For an infinitesimal scale transformation $\lambda=1+\epsilon$, we have 
\aln{\Xi_{1+\epsilon}^{}[T,V;U(x)]=\Xi_{}^{}[T,(1+(d-1)\epsilon)V;U(x(1-\epsilon))-(d-1)T\epsilon]~,
}
or equivalently the grand potential satisfies 
\aln{
-\beta W_{1+\epsilon}^{}[T,V;U(x)]
=-\beta W^{}[T,(1+(d-1)\epsilon)V;U(x(1-\epsilon))-(d-1)T\epsilon]~. 
\label{eq:relation1}
}
The Legendre transformation of the left hand side gives $-\beta \Gamma_{1+\epsilon}^{}[T,V;\rho(x)]$ by definition. 
On the other hand, the Legendre transformation of the right hand side is
\aln{
&\underset{U}{\text{Min}}\left[-\beta W^{}[T,(1+(d-1)\epsilon)V;U(x(1-\epsilon))-(d-1)T\epsilon]-\beta\int_{V}d^{d-1}xU(x)\rho(x)
\right] 
\nn
=& \underset{U}{\text{Min}}\bigg[-\beta W^{}[T,(1+(d-1)\epsilon)V;U(x(1-\epsilon))-(d-1)T\epsilon]  \nn
& -(1-(d-1)\epsilon)\beta\int_{(1+(d-1)\epsilon)V}d^{d-1}x\left\{U(x(1-\epsilon))-(d-1)T\epsilon\right\}\rho(x(1-\epsilon)) 
\bigg]
-(d-1)\epsilon\int_V d^dx\rho(x)
\nn
=&-\beta \Gamma^{}[T,(1+(d-1)\epsilon)V;(1-(d-1)\epsilon)\rho(x(1-\epsilon))]-(d-1)\epsilon\int_V^{} d^{d-1}x\rho(x)~,
}
%
which leads to
\aln{
-\beta \Gamma_{1+\epsilon }^{}[T,V;\rho(x)]=-\beta \Gamma^{}[T,(1+(d-1)\epsilon)V;(1-(d-1)\epsilon)\rho(x(1-\epsilon))]-(d-1)\epsilon\int_V^{} d^{d-1}x\rho(x)+{\cal{O}}(\epsilon^2)~.
}
This relation corresponds to Eq.~(\ref{general compensation}) in the general formalism.    
By taking the functional derivative with respect to $\rho(x)$, we have\footnote{By writing the parameter dependences explicitly, the $l$-th order functional derivative of the first term in the R.H.S is calculated as 
\aln{
&(1-(d-1)l\epsilon)\left(\prod_{i=1}^l\int d^{d-1}y_i^{}
\frac{\delta (\rho(y_i^{})-\epsilon y_i^\mu\partial_\mu^{}\rho(y_i^{}))}{\delta \rho(x_i^{})}\right)c_l^{}(\{y_i^{}\},T,V+(d-1)\epsilon V;\rho)
\nn
&=(1-(d-1)l\epsilon)c^{(l)}(\{x_i^{}\},T,V+(d-1)\epsilon V;\rho)
+\epsilon\sum_{i=1}^l \partial_{i\mu}^{}(x_i^\mu
c^{(l)}(\{x_i^{}\},T,V;\rho))
 \nn
 &=c^{(l)}(\{x_1^{}\},T,V;\rho)+(d-1)\epsilon \frac{\partial}{\partial\log V}c^{(l)}(\{x_1^{}\},T,V;\rho)
+\epsilon \sum_{i=1}^l x_i^\mu
 \partial_{i\mu}^{}c^{(l)}(\{x_1^{}\},T,V;\rho)
}
}
\aln{
c_{1+\epsilon}^{(l)}(\{x_i^{}\})=&c^{(l)}(\{x_1^{}\})+(d-1)\epsilon \frac{\partial}{\partial\log V}\bigg|_{T,N}^{}c^{(l)}(\{x_i^{}\})
\nn
&+\epsilon \sum_{i=1}^l x^\mu \partial_\mu^{} c_{}^{(l)}(\{x_i^{}\})-\epsilon \delta_{l0}(d-1)\int_V d^{d-1}x\rho(x)- \epsilon \delta_{l1}(d-1)~.
}
\aln{
\therefore\quad \frac{dc_{1+\epsilon}^{(l)}(\{x_i^{}\})}{d\epsilon}\bigg|_{\epsilon=0}&=\left[(d-1) \frac{\partial}{\partial\log V}\bigg|_{T,N}^{}+\sum_{i=1}^l x_i^\mu \partial_{i\mu}^{}\right]c_{}^{(l)}(\{x_i^{}\})
\nn
&\hspace{3cm} -\delta_{l0}(d-1)\int_V d^{d-1}x\rho(x)- \delta_{l1}(d-1)~.
}
On the other hand, the left hand side can be also calculable by taking the functional derivatives of the flow equation~(\ref{flow in scale}). 
Thus, we obtain 
\aln{
&\left[(d-1) \frac{\partial}{\partial\log V}\bigg|_{T,N}^{}+\sum_{i=1}^l x_i^\mu \partial_{i\mu}^{}\right]c_{}^{(l)}(\{x_i^{}\})-\delta_{l0}(d-1)\int_V d^{d-1}x\rho(x)- \delta_{l1}(d-1)
\nn
=&-\frac{\beta}{2}\int d^{d-1}x\int d^{d-1}y~
 \frac{\delta^l n^{(2)}(x,y)}
 {\delta \rho(x_1^{})\cdots \delta \rho(x_l^{})}(x-y)^i\partial_i^{}v(x-y)~. 
\label{DRG}
}
%
Since the volume derivative is related to the density derivative by
\aln{\frac{\partial}{\partial \log V}\bigg|_{T,N}^{}=-\frac{\partial}{\partial \log n}\bigg|_{T,N}^{}~,
}
Eq.~(\ref{DRG}) describes the response of the system to the  density variation.  
See also Ref.~\cite{Iso:2018rag} for the calculations of the right hand side in Eq.~(\ref{DRG}). 
Note that we used the field theoretical approach in the previous work~\cite{Iso:2018rag} to derive the above flow equation.  
However, one can see that it is merely a direct consequence of general response theory.  
%

\subsection{Classical nonequilibirum systems}\label{nonequilibirum systems}
Response theory developed in Section~\ref{sec:response} can be also applied to classical nonequilibirum systems because the latter can be cast into the path integral formulation in some systems~\cite{PhysRevA.8.423,Janssen,deDominicis}. 
Here, we focus on the Langevin stochastic dynamics. 
In the following, $x$ denotes a spacetime point $x\in \Sigma_{d}^{}$ while $t$ is the Langevin time.    

We consider a real scalar field $\phi(t,x)$ which obeys the Langevin equation
\aln{\partial_t^{}\phi(t,x)=-F[\phi(t,x)]+\xi(t,x)~,
\label{Langevin system}
}
where $F[\phi(t,x)]$ is a general external force 
and $\xi(t,x)$ denotes the Gaussian random force i.e. its  probability distribution is given by
\aln{
&P[\xi(t,x)]={\cal N}\exp\left(-\frac{1}{2}\langle \xi|G^{-1}|\xi\rangle\right)~,\label{distribution of xi}
}
where
\aln{
\langle \xi|G^{-1}|\xi\rangle=\int dt\int dt'\int d^{d}x\int d^{d}x'\xi(t,x)G^{-1}(t,x;t',x')\xi(t',x')~.
} 
Here ${\cal N}$ is a normalization factor. 
Correspondingly, the correlation function of $\xi(t,x)$ is given by 
\aln{\langle \xi(t,x)\xi(t',x')\rangle=G(t,x;t',x')~.
}
The Langevin system (\ref{Langevin system}) can be cast into a field theory based on the Martin-Siggia-Rose-Janssen-de Dominicis formalism~\cite{PhysRevA.8.423,Janssen,deDominicis} as follows. 
The expectation value of an operator ${\cal O}[\phi(t,x)]$ can be written as 
\aln{
\langle {\cal O}[\phi(t,x)]\rangle &=\left\langle \int {\cal D}\phi {\cal O}[\phi(t,x)]\prod_{t,x}\delta\left(\partial_t^{}\phi(t,x)-F[\phi(x,t)]-\xi(t,x)\right) 
 \right\rangle
 \\
 &\propto \left\langle \int {\cal D}\phi \int {\cal D}\phi' {\cal O}[\phi(t,x)]\exp\left(i\int dt\int d^{d}x\phi'(t,x)\left\{\partial_t^{}\phi(t,x)+F[\phi(x,t)]-\xi(t,x)\right\}\right) 
 \right\rangle 
 \\
 &=\int {\cal D}\phi \int {\cal D}\phi' {\cal O}[\phi(t,x)]
e^{i\int dt\int d^{d}x\phi'(t,x)\left(\partial_t^{}\phi(t,x)+F[\phi(x,t)]\right)}
\nn
&\hspace{3cm}\times \left\langle\exp\left(-i\int dt\int d^{d}x\phi'(t,x)\xi(t,x)\right) 
 \right\rangle ~,
}
where we have introduced another real scalar $\phi'(t,x)$ and 
$\langle \cdots\rangle$ is the expectation value by the distribution (\ref{distribution of xi}). 
By using Eq.~(\ref{distribution of xi}), the above expectation value is calculated as 
\aln{\left\langle\exp\left(-i\int dt\int d^{d}x\phi'(t,x)\xi(t,x)\right) 
 \right\rangle&={\cal N}\int {\cal D}\xi \exp\left(-\frac{1}{2}\langle \xi|G^{-1}|\xi\rangle-i\langle \phi'|\xi\rangle \right)
 \nn
 &=\exp\left(\frac{1}{2}\langle \phi'|G|\phi'\rangle\right)~,
}
which leads to the following path integral expression
\aln{
\langle {\cal O}[\phi(t,x)]\rangle=\frac{1}{Z}\int {\cal D}\phi\int {\cal D}\phi'{\cal O}[\phi(t,x)]
e^{iS[\phi,\phi']}
}
where the action is given by~\footnote{In the literatures, $\overline{\phi}:=i\phi'$ is often regarded as an elementary field. 
}
\aln{S[\phi,\phi']=\int dt\int d^{d}x\phi'(t,x)\left\{\partial_t^{}\phi(t,x)+F[\phi(x,t)]\right\}+\frac{1}{2i}\langle \phi'|G|\phi'\rangle~.
\label{action of nonequilibirum}
%
}
One can see that the effect of random force $\xi(t,x)$ is now taken placed by another real scalar field $\phi'(t,x)$ and the system is now described by a two-scalar QFT in a $(d+1)$ dimensional spacetime. 
%
%
As in the FRG, we can add a regulator function:
\aln{\Delta S=-\frac{1}{2}\sum_{i,j=1}^2\langle \phi_i^{}|R_{k,ij}^{}|\phi_j^{}\rangle~,
} 
where $\phi_{1(2)}^{}=\phi~(\phi')$ and the regulator function $R_{k,ij}^{}(x-y)$ is now a $2\times 2$ matrix. 
Then, by performing the same calculations as  Section~\ref{sec:FRG}, we obtain 
\aln{
\label{flow equation in classical nonequilibirum}
\frac{d(i\Gamma_k^{}[\phi,\phi'])}{dk}=-\frac{1}{2}{\rm Tr}\left((\partial_k^{}R_{k,ij}^{})(\Gamma_{k,ji}^{(2)}-R_{k,ji}^{})^{-1}
\right)~,
} 
where 
\aln{
\Gamma_{k,ji}^{(2)}=\frac{\delta \Gamma_k^{}[\phi,\phi']}{\delta \phi_i^{}\delta \phi_j^{}}~.
}
Note that the trace in Eq.~(\ref{flow equation in classical nonequilibirum}) is taken over the $d+1$ dimensional functional space. 
The generalization to a vector field $\phi_a^{}(t,x)~(i=1,2,\cdots,N)$ is straightforward. 
In particular, one can consider the velocity fields $\phi_\mu^{}(t,x)=v_\mu^{}(t,x)~(\mu=0,1,\cdots,d-1)$ which obey the Navier-Stokes equation:  
\aln{
\partial_t^{}v_\mu^{}(t,x)&=-v^\nu \partial_\nu^{}v_\mu^{}(t,x)-\frac{1}{\rho(t,x)}\partial_\mu^{}p(t,x)+\cdots +\xi_\mu^{}(t,x)
\\
&:=-F[\{v_\mu^{}(t,x)\}]+\xi_\mu^{}(t,x)~,
} 
where the correlation function of the random force is now written as 
\aln{\langle \xi_\mu^{}(t,x)\xi_\nu^{}(t',x')\rangle=G_{\mu\nu}^{}(t,x;t'x')~.
}
Then, by following the MSRJD formalism, the field theoretic  action is given by
\aln{
S[\{v_\mu^{}\},\{v_\mu^{'}\}]=\int dt \int d^dx\left\{
v_\mu^{'}\left(\partial_t^{}v^\mu+F[v]\right)
\right\}+\frac{1}{2i}\langle v'|G|v'\rangle~.
}
By adding a regulator 
\aln{\Delta S=-\frac{1}{2}\sum_{\mu,\nu=0}^{d-1}\langle v_\mu^{}|R_{k}^{\mu\nu}|v_\nu^{}\rangle~,
}
we now obtain the exact flow equation for the velocity fields:
\aln{\frac{d(i\Gamma_k^{}[\{v_\mu^{}\},\{v'_\mu\}])}{dk}=-\frac{1}{2}{\rm Tr}\left((\partial_k^{}R_k^{\mu\nu}){{(\Gamma_k^{(2)}}^{\nu\mu}-R_k^{\nu\mu})^{-1}}
\right)~,
}
which is again the same form as the Wetterich's equation. 
The functional RG method has proven to be very powerful to understand the various properties of fluid systems such as turbulence~\cite{PhysRevE.93.063101,TOMASSINI1997117,PhysRevE.86.016315} and long-range behaviours of correlation functions~\cite{PhysRevE.95.023107,Tarpin_2018,Tarpin_2019}. 
See also Ref.~\cite{Dupuis:2020fhh} for a recent review. 


\section{Summary and discussion}\label{sec:summary}
In this paper, we have discussed the basic ideas and formulations of functional flow approach in equilibrium systems. 
%
%
By considering the response of equilibrium systems to a general variation of the parameters, we obtained functional flow equations of the generating functionals. 
%
Conventionally, some IR cut-off parameter is often introduced in the two-body interaction $v_2^{}(x,y)$, which leads to the Wetterich type flow equations. 
However, this is not the only and unique way to introduce an artificial parameter and one can also consider similar procedures 
even for the higher-order potential terms $v_n^{}(\{x_i^{}\})~(n\geq 3)$ in general.  
Such a treatment would be useful when a system exhibits long-range correlations, and our response theory provides a straight and systematic way to obtain the functional flow equations for such general cases.      
In this paper, we have focused on explaining the theoretical bases of response theory and not studied any concrete systems/models.   
We would like to investigate them in future publications.  
%
%

\section*{Acknowledgements} 
We would like to thank Satoshi Iso, Yoshimasa Hidaka, Sinya Aoki, Kengo Shimada and Philip Lu for the valuable discussions and comments.   
%

\appendix

\bibliographystyle{TitleAndArxiv}
\bibliography{Bibliography}

\end{document}